\documentstyle[11pt,epsfig,newpasp,twoside]{article}
\def\edcomment#1{\iffalse\marginpar{\raggedright\sl#1\/}\else\relax\fi}
\reversemarginpar

\def\apj#1{{\em Astrophys. J.} {\bf #1}}
\def\aj#1{{\em Astron. J.} {\bf #1}}
\def\mnras#1{{\em Mon. Not. R. astr. Soc.} {\bf #1}}
\def\aap#1{{\em Astron. Astrophys.} {\bf #1}}
\def\nat#1{{\em Nature} {\bf #1}}
\def\apjs#1{{\em Astrophys. J. Suppl.} {\bf #1}}
\def\prl#1{{\it Phys. Rev. Lett.} {\bf #1}}
\def\araa#1{{\it Ann. Rev. Astron. Astrophys.} {\bf #1}}
\def\azh#1{{\it Astron. Zh.} {\bf #1}}
\def\apss#1{{\it Astrophys. Space Sc.} {\bf #1}}
\def\pasp#1{{\it Publ. ASP} {\bf #1}}
\def\etal{{et al.\/}\ }
\def\Mpc{$h^{-1}$~{\rm  Mpc}}

\begin{document}

\title{Dark Matter and Large Scale Structure}

\author{Jaan Einasto}

\affil{ Tartu Observatory, EE-61602 T\~oravere, Estonia}

\begin{abstract}
A review of the study of dark matter and large scale structure of
the Universe at Tartu Observatory is given.  Tartu astronomers have
participated in this development, starting from Ernst \"Opik and
Grigori Kuzmin, and continuing with the present generation of
astronomers. Our goal was to understand better the structure, origin
and evolution of the Universe.
\end{abstract}

\section{Prologue}

Once I happened to read the Thomas Kuhn (1970) book {\em The Structure of
Scientific Revolutions}.  The presence of dark matter around galaxies had
just been reported, and the fight between supporters and opponents of the
dark matter concept was at its peak.  Then I realised that the dark matter
story seems to be a good example of a scientific revolution.  Ten years
later, in concluding remarks of the IAU Symposium on ``Dark Matter in the
Universe'', Scott Tremaine also pointed to the development of the dark
matter concept as a classical example of a scientific revolution (Tremaine
1987, see also Binney \& Tremaine 1987).

Tartu astronomers have participated in the dark matter study for a long
time, starting from Ernst \"Opik -- the founder of the contemporary
astronomy school in Estonia; and followed by Grigori Kuzmin, the most
talented student of Ernst \"Opik.  The present generation of
astronomers has continued the study of dark matter. Unexpectedly,
this work has lead us to study the distribution of galaxies and
clusters on large scales and led to the discovery of voids and filaments.

In the following I shall give a review of the study of dark matter and
large scale structure of the Universe, as seen by an Estonian
astronomer.  This story is biased, and is to be complemented by
similar stories of the development of the dark matter concept
elsewhere.  However, I hope that our story is of interest to the
cosmology community, as some aspects of it are not well documented so
far.

\section{Local Dark Matter}

\subsection{Ernst \"Opik}

An astrophysical approach in astronomy was introduced at Tartu
Observatory with the work by Ernst \"Opik.  He started his studies
being a student of Moscow University.  One of the first questions he
was curious about was the possible presence of invisible matter in our
Galaxy.  He developed a method to determine the density of matter near
the Galactic plane using vertical oscillations of stars.  He concluded
that there is no evidence for large amounts of invisible matter near
the Galactic plane (\"Opik 1915).

\begin{figure}[ht]
\begin{center}
\begin{minipage}{0.7\textwidth}
\plotone{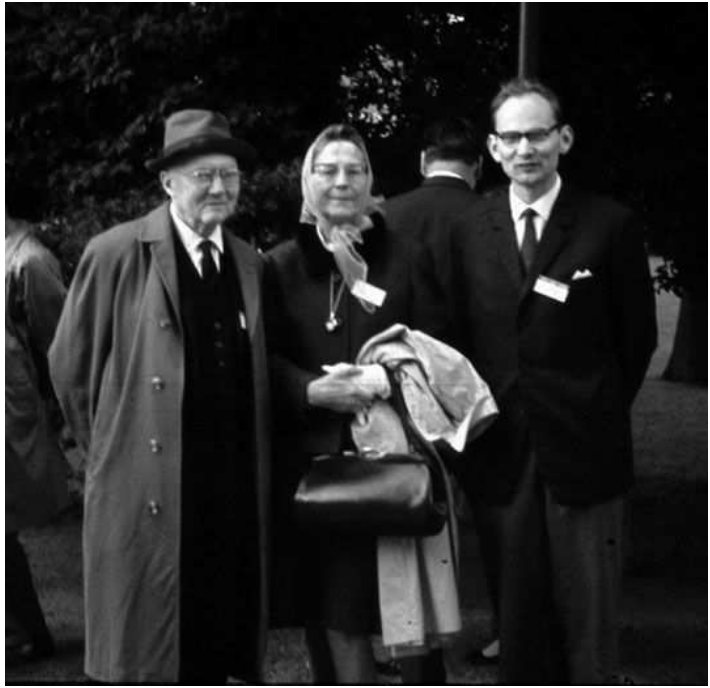}
\end{minipage}
\end{center}
\caption{Ernst \"Opik with his wife Alice and author in Brighton
during the IAU General Assembly 1970.}
\label{fig:Opik}
\end{figure}

A hot topic was the discussion on the nature of spiral nebulae: Are
they gaseous objects within the Milky Way or distant worlds similar in
structure to our Galaxy?  \"Opik was also interested in this
problem. In 1918 he delivered a talk at the Meeting of the Moscow
Society of Amateur Astronomers, devoted to his study of the structure
of the Andromeda Nebula, M31.  Shortly before, the first relative
velocity measurements of the central bulge of M31 had been published,
and \"Opik had developed a method for estimating distances to spiral
nebulae from relative velocities within them. He used Newton's law of
gravity, and the apparent luminosity of the central part of M31
derived from photometric observations.  Further he assumed that the
mass--to--luminosity ratio in M31 is identical to the ratio in our
Galaxy, and developed a mass distribution model of M31 to estimate a
parameter in the formula for relative velocities. He accepted for the
mass-to-luminosity ratio in the Solar vicinity a value 2.63 in Solar
units (Kapteyn 1904, van Rhijn 1920), and \"Opik's (1915) own estimate
of the mass density of the matter in the Solar neighbourhood.  From
these data he got for the distance of the Andromeda Nebula 440 kpc
(\"Opik 1922).  With this work he solved the problem of the nature of
spiral nebulae, and developed mass distribution models of galaxies --
a prerequisite for further work which culminated 50 years later with
the discovery of dark halos of galaxies.

\subsection{Grigori Kuzmin}

The work on galactic mass modelling was continued by Grigori Kuzmin.
In the 1940s he developed a new method for galactic mass modeling.  He
applied his theory first to the Andromeda galaxy using recently
published rotation data by Babcock (1939).  Due to war time he
published his results only in Estonian in the yearbook of the Tartu
Observatory (Kuzmin 1943).

\begin{figure}[ht]
\begin{center}
\begin{minipage}{0.7\textwidth}
\plotone{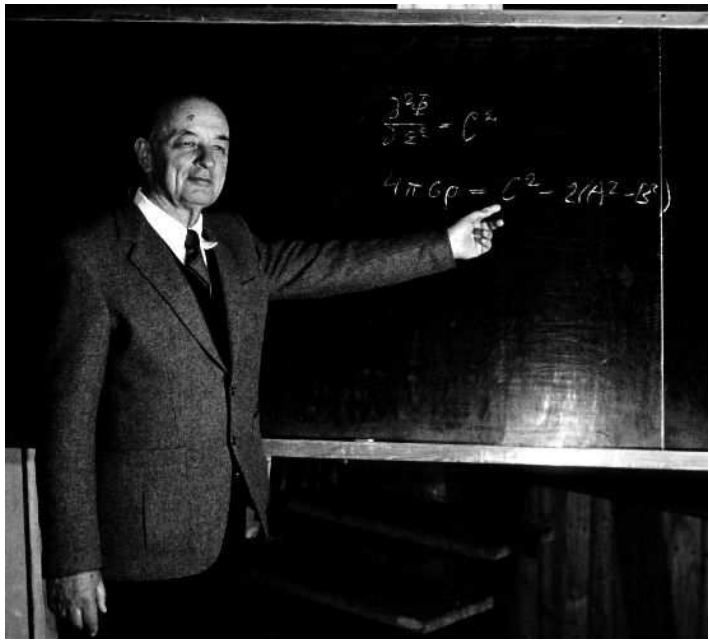}
\end{minipage}
\end{center}
\caption{Grigori Kuzmin explaining his formula to calculate the
dynamical density of matter in the Galaxy, late 1970s.}
\label{fig:Kuzmin}
\end{figure}

Next Kuzmin turned his attention to our own Galaxy.  Here the central
problem was the density of matter in the Solar vicinity.  The mass
density is given by the Poisson equation.  Kuzmin expressed the
derivative of the gravitational potential in the Galactic plane
through Oort constants, $A$ and $B$, and the derivative of the
gravitational potential in the vertical direction, $\partial
\Phi/\partial z$, through a constant $C$, which characterises the
vertical gravitational acceleration; it has the same dimension as the
$A$ and $B$, and is called the Kuzmin constant.  This derivative can
be expressed through the ratio of dispersions of velocities and
coordinates in vertical direction, $C=\sigma_z/\zeta_z$ (Kuzmin 1952a,
1955).  Finally the Poisson equation takes the form:
\begin{equation}
4\pi G \rho = C^2 - 2(A^2 - B^2).
\label{eq:pois2}
\end{equation}
Kuzmin (1952a, 1955) used data on the distribution of A and gK stars
and analysed results obtained in earlier studies by Oort (1932) and
others.  He got a weighted mean value $C=
68$~km\,s$^{-1}$\,kpc$^{-1}$, which leads to the density estimate
$\rho = 0.08$~$M_{sun}$\ pc$^{-3}$, in good agreement with direct
density estimates of all known stellar populations (including
estimates for the mass in invisible low--mass stars and white
dwarfs). Two students of Kuzmin made independent analyses, using
different methods and initial data (Eelsalu 1959, J\~oeveer 1972,
1974b) and confirmed Kuzmin results.  Thus we came to the conclusion
that {\em there is no evidence for the presence of dark matter in the
disk of the Galaxy}.

In parallel to the analysis of the spatial density Kuzmin developed a
detailed model of the Galaxy (Kuzmin 1952b, 1953, 1956a, b). Here, for
the first time, non-homogeneous ellipsoids were used instead of sums
of ellipsoids of constant density; also the third integral of motion
of stars was applied, and principles of calculation of 3--dimensional
stellar orbits were presented.  The mass model of the Galaxy by
Schmidt (1956) still uses sums of ellipsoids of constant density; also
a higher value of the local density was accepted.  The local density
problem was studied again by Hill (1960) and Oort (1960); both
obtained considerably higher local densities of the matter, and argued
that there exist large amounts of dark matter in the Galactic disk.
The discrepancy between Tartu and Dutch results for the presence or
absence of the local dark matter in the disk of the Galaxy was not
solved until recently.  Modern data have confirmed the results by Kuzmin
and his collaborators (Gilmore, Wyse \& Kuijken 1989).

\section{Global Dark Matter}

\subsection{Galactic Models}

Problems of the structure and evolution of stars and stellar systems were
a central issue at Tartu Observatory.  \"Opik (1938) developed the modern
theory of stellar evolution based on the burning of hydrogen in stellar
cores, which leads to the formation of red giant stars after the main
sequence stage.  Rootsm\"ae (1961) applied these ideas to kinematics of
stars to find the sequence of formation of different stellar populations;
similar ideas were developed independently by Eggen, Lynden--Bell and
Sandage (1962). Kuzmin developed the major principles of galactic
modeling.  Thus, in the 1950s we had a concept of the structure and
evolution of stellar populations in galaxies which is rather close to the
presently accepted picture.  It was quite natural to apply this knowledge
to construct more detailed models of galaxies, including explicitly main
stellar populations.

My first experience of galaxy modeling was in 1952 when I made
calculations for the Kuzmin new model of the Galaxy.  At this time a
popular model was presented by a leading Soviet astronomer Parenago
(1950).  In the Parenago model a circular velocity law was accepted
which has a rather fast decrease on the periphery and led to
negative spatial densities.  I suggested an improvement of the model:
we applied the velocity law only until the galactocentric distance
where the density becomes zero; and we assumed zero density outside this
limit. At larger distances the circular velocity was calculated from
the density law.

I made my PhD thesis on kinematics of stars and returned to galactic
modeling again in the 1960s.  The modeling proceeded step by step.
Detailed local structure is known only for our own Galaxy, and global
information on stellar populations is better known for external galaxies.  
Thus it was natural to study our Galaxy and our next neighbour, the
Andromeda galaxy, in parallel.  The first step was to find a balanced
system of Galactic constants, using all available data. This system was
presented in the Commission 33 Meeting of the 12th General Assembly of IAU
(Einasto and Kutuzov 1964); it can be considered as the first version of
the presently accepted IAU system of Galactic constants.  A composite
model of the Galaxy based on this system of constants was calculated
(Einasto 1965).  Then I realised that there was a need for a more detailed
method for the construction of composite models of galaxies.  This goal
was realised in a series of papers in Tartu Publications, and a summary
was published in a journal (Einasto 1969a).  The main principles of model
construction were: galaxies can be considered as sums of physically
homogeneous populations; physical properties of populations
(mass--to--luminosity ratio, colour) should be in agreement with models of
physical evolution of stellar populations; the density of a population can
be expressed as ellipsoids of constant flatness and rotational symmetry;
densities of populations are non--negative and finite; moments of
densities which define the total mass and effective radius of the galaxy
are finite.  It was found that in a good approximation densities of all
stellar populations can be expressed by a generalised exponential law:
$\rho (a)=\rho (0)\exp [-(a/a_c)^{1/N}]$, where $\rho (0)=hM/(4\pi\epsilon
a_0^3)$ is the central density, $a= \sqrt{R^2+z^2/\epsilon^2}$ is the
distance along the major axis, $\epsilon$ is the axial ratio of the
equidensity ellipsoid,  $a_c=ka_0$ is the core radius ($a_0$ is the
harmonic mean radius), $h$ and $k$ are normalizing parameters, depending
on the structural parameter $N$, which allows to vary the density
behaviour with $a$. The definition of normalizing parameters and their
calculation was described by Einasto (1970b). The cases $N=1$ and $N=4$
correspond to conventional exponential and de Vaucouleurs models,
respectively.

The method was applied for the Andromeda galaxy (Einasto 1969b, 1970a,
Einasto \& R\"ummel 1970a), and for our Galaxy (Einasto 1970b).  In
the case of the Andromeda galaxy we encountered two problems.  The
first problem concerns rotation data and physical properties near the
centre of the galaxy.  If we accepted rotational velocities, based
mostly on radio observations (Roberts 1966), then the
mass--to--luminosity ratio, $M/L$, of central stellar populations
became very low, of the order of 1 in Solar units. On the other hand,
spectral data (Spinrad 1966) suggested a much higher value, $M/L
\approx 17$. To solve this discrepancy, we analysed the velocity field
obtained from radio observations, and demonstrated that low rotational
velocities in central regions are due to low spatial resolution of the
radio beam (Einasto \& R\"ummel 1970b,c).  The corrected velocity
field was in agreement with a higher value of $M/L$ in central regions
of M31.  As a further test the rate of star formation as function of the
density of stellar populations was studied in M31, and the results
confirmed the Schmidt (1959) law: the star formation rate is proportional
to the density squared (Einasto 1972). 

Additionally, a model of physical evolution of stellar populations was
developed (Einasto 1971).  The model was similar to the model by
Tinsley (1968), but some aspects were developed in more detail.  The
model used as input data the evolutionary tracks of stars of various
composition (metallicity) and age; the star formation rate from
Salpeter (1955) as a function of stellar mass was used, as well as a
low--mass limit of star formation of $M_0 = 0.03~M_{sun}$. The model
yielded a continuous sequence of population parameters as a function
of age (colour, spectral energy distribution, $M/L$).  The results of
modeling stellar populations were compared with direct dynamical data
for central regions of galaxies (velocity dispersions) by Einasto \&
Kaasik (1973).  These data supported relatively high values
($M/L\approx 10 - 30$) for old metal--rich stellar populations near
centres of galaxies; moderate values ($M/L\approx 3 - 10$) for discs
and bulges; and low values ($M/L\approx 1 - 3$) for metal--poor
halo--type populations.  Modern data yield slightly lower values, due
to more accurate measurements of velocity dispersions in central
regions of galaxies.

The second problem encountered in the modeling of M31 was the rotation
and density distribution on the periphery.  If rotation data were
taken at face value, then it was impossible to represent the
rotational velocity with the sum of known stellar populations. The local
value of $M/L$ increases towards the periphery of M31 very rapidly if the
mass distribution is calculated directly from rotation velocity.  All
known old metal--poor halo--type stellar populations have a low $M/L
\approx 1$; in contrast on the basis of rotation data we got $M/L >
1000$ on the periphery of the galaxy near the last point with measured
rotational velocity.

There were two possibilities to solve this problem: to accept the
presence of a new population with very uncommon properties, or to
assume that on the periphery of galaxies there exist non--circular
motions.  We found that the first alternative has several serious
difficulties.  If the hypothetical population is of stellar origin, it
must be formed much earlier than known populations, because all known
stellar populations form a continuous sequence of kinematical and
physical properties (Rootsm\"ae 1961, Einasto 1974a), and there is no
place where to put this new population into this sequence.  Secondly,
the star formation rate is proportional to the square of the local
density (Schmidt 1959, Einasto 1972), thus stars of this population
should have been formed during the contraction phase of the formation
of the population near its central more dense regions, and later
expanded to the present distance.  The only source of energy for
expansion is the contraction of other stellar populations. The
estimated total mass of the new population exceeded the summed mass of
all previously known populations.  Estimates of the energy needed for
the expansion demonstrated that the mass of the new population is so
large that even the contraction of all other stellar populations to
zero radius would not be sufficient to expand the new population to
its present size.  And, finally, it is known that the star formation
is not an efficient process (usually in a contracting gas cloud only
about 1~\%~ of the mass is converted to stars); thus we have a problem
how to convert, in an early stage of the evolution of the Universe, a
high fraction of primordial gas to this population of stars.  Taking
into account all these difficulties we accepted the second alternative
-- the presence on non--circular motions (Einasto 1969b), similar to
many other astronomers (see Materne \& Tammann 1976).  As we soon
realised, this was a wrong decision.

\subsection{Galactic Coronae}

In spring 1972 I was asked to give an invited review on Galactic
models at the First European Astronomy Meeting in Athens.  At this
time population models of galaxies had been calculated already for 5
galaxies of the Local Group and the giant elliptical galaxy M87 in the
Virgo cluster.  More and more data accumulated on rotation velocities
of galaxies.  New data suggested the presence of almost flat rotation
curves on the periphery of galaxies, thus it was increasingly
difficult to accept the previous concept of large non--circular
motions.  On the other hand, recently finished calculations of the
physical evolution of stellar populations confirmed our previous view
that it is extremely difficult to accept a stellar origin of the
hypothetical population.  In summer 1972 I discussed the problem with
my collaborator Enn Saar.  He suggested to abandon the idea that only
stellar populations exist in galaxies, to assume that there is a
population of unknown nature and origin and to look which properties
it should have using available data on known stellar populations.
Here we can say with the words of Sherlock Holmes {\em ``When you have
eliminated the impossible, whatever remains, however improbable, must
be the truth''} (cited by Binney \& Tremaine 1987).  Quickly a second
set of models for galaxies was calculated, and parameters for the new
dark population were found (for the distribution of mass-to-luminosity
ratio in galaxies see Figure~3).  My talk in Athens was on September 8,
1972. The main results were (Einasto 1974b): (1) There are two dark
matter problems: the local and the global one; (2) The local dark
matter, if it exists, must be of stellar origin, as it is strongly
concentrated to the Galactic plane; (3) The global dark matter is of
non--stellar origin; it has very low concentration to the plane and
centre of the galaxy; its dynamical and physical properties are
different from all previously known stellar populations.  To avoid
confusion with the conventional halo population I suggested to call
the new population ``corona''; (4) Available data are insufficient to
determine outer radii and masses of coronae. Preliminary estimates
indicated that in some galaxies the mass and radius of the corona may
exceed considerably the mass and radius of stellar populations.

\begin{figure}[ht]
\begin{center}
\begin{minipage}[b]{0.40\textwidth}
\plotone{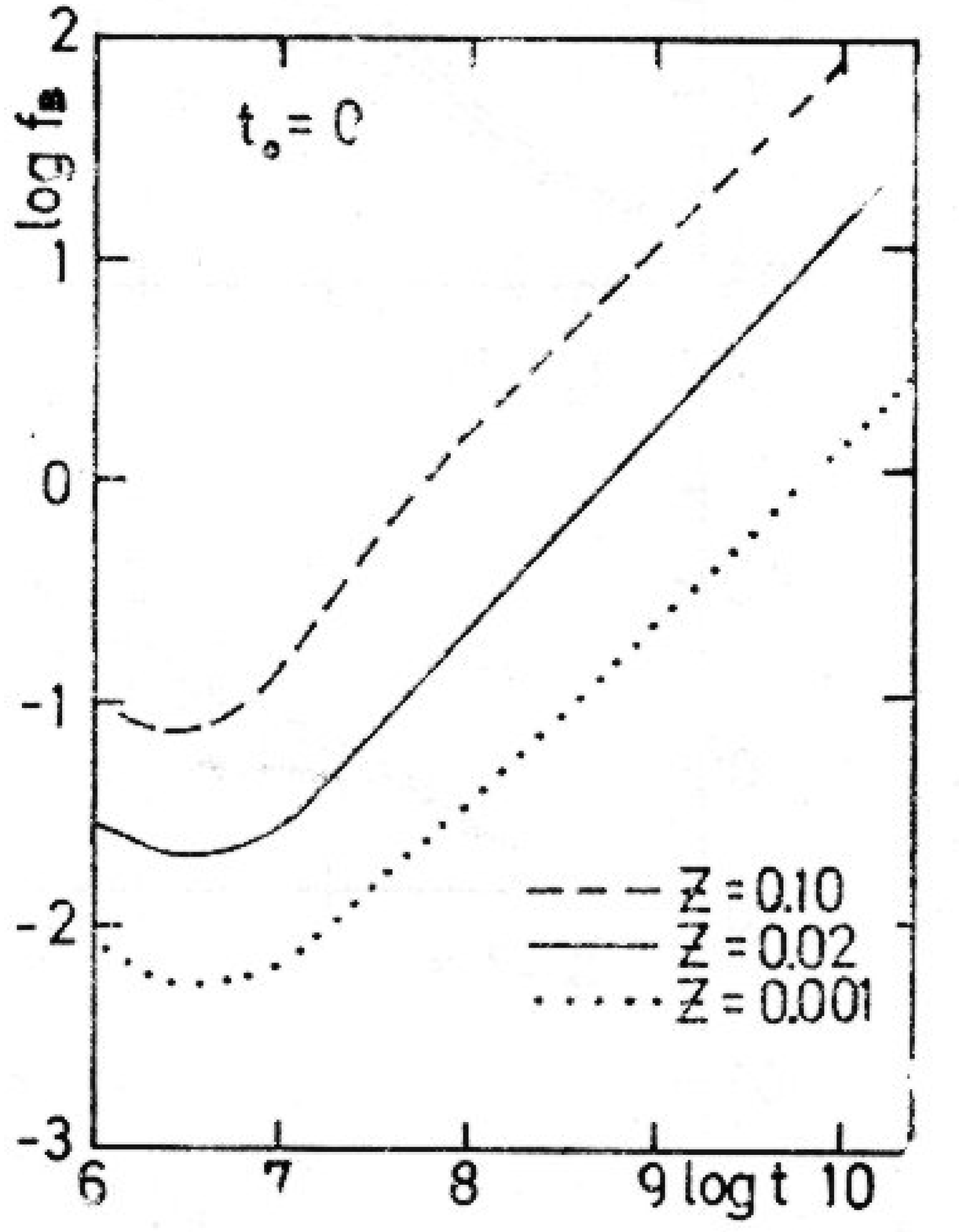}
\end{minipage}
\begin{minipage}[b]{0.56\textwidth}
\plotone{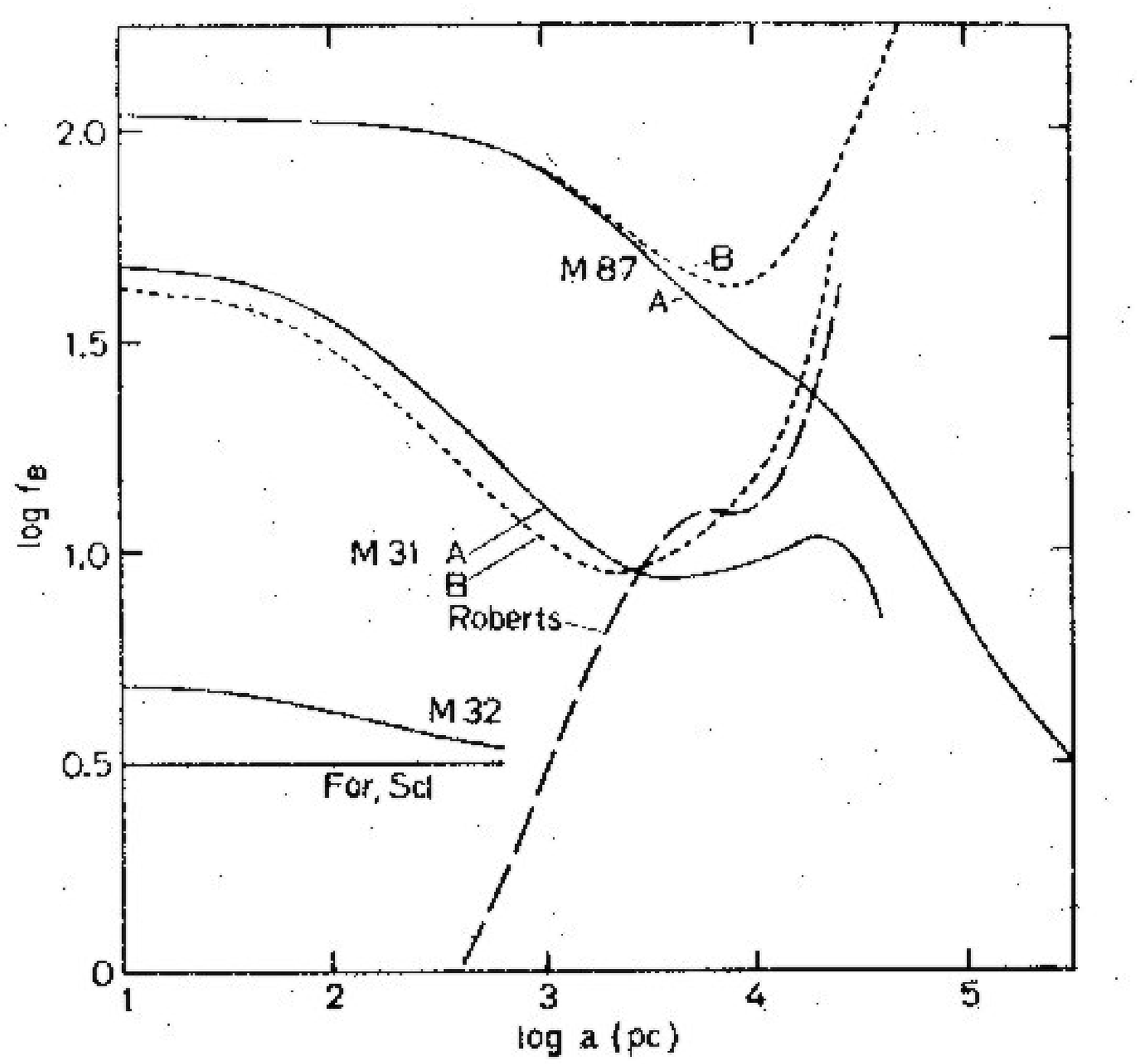}
\end{minipage}
\end{center}
\caption{Left: evolution of the
mass-to-luminosity ratio $f_B$ for stellar populations of different
metallicity $Z$ and instant star formation (Einasto 1971). Age $t$ is
given in years. Right: Distribution of mass-to-luminosity ratio, $f_B=M/L_B$,
in galaxies of the Local Group and M87: models without (A) and with
(B) dark corona (Einasto 1974b). }
\label{fig:dm}
\end{figure}

The Athens report did not give rise to special excitement.  The main
reason for this lukewarm reception was probably the absence of a solid
proof for the existence of the corona, of its main parameters (mass
and radius), and of its nature.  Thus I continued the search for
further evidence.  Soon I noticed that the problem of galactic coronae
is the same as discussed already long time ago in clusters and groups
of galaxies, starting from the pioneering work by Zwicky (1933) and
Kahn and Woltjer (1959).  The problem was discussed in detail at the
Santa Barbara Conference on the Instability of Systems of Galaxies
(Neyman, Page \& Scott 1961).  I fully agreed with arguments by van
den Bergh (1961) that clusters of galaxies are old systems, thus they
must be stabilised by large masses.  A similar problem exists in
double elliptical galaxies. The mean mass--to--luminosity ratio of
double elliptical galaxies $M/L\approx 66$ (Page 1952, 1960), whereas,
according to our recent population models, stellar populations had $1
\leq M/L \leq 30$.  In particular, for bulges of galaxies our
composite galactic model gave $M/L\approx 3 - 10$ in good agreement
with models of physical evolution of stellar populations (see
Figure~3).  Low estimates for $M/L$ in bulges were confirmed by Faber
\etal (1977) who measured rotation velocities near the edge of the
bulge of the Sombrero galaxy and found $M/L\approx 3$.  Thus it was
evident that elliptical galaxies should have massive coronae.

Reading these papers on the mass discrepancy in clusters and
elliptical galaxies I realised how it is possible to check the
presence of dark coronae around galaxies.  If coronae are large
enough, then in pairs of galaxies the companion galaxy can be
considered as a test particle to measure the gravitational attraction
of the main galaxy.  Mean relative velocities, calculated for
different distances from the main galaxy, can be used instead of
rotation velocities to find the mass distribution of giant galaxies.
Quickly I collected data for pairs of galaxies, and on January 11,
1974 I had the results: radii and masses of galactic coronae exceeded
radii and masses of parent galaxies by an order of magnitude!
Together with A. Kaasik and E. Saar we calculated new models of
galaxies including dark coronae.

In those years Soviet astronomers had the tradition to gather in Caucasus
Winter Schools. Results of galactic mass modeling were reported in a
Winter School in 1972. That year the School was hold near the Elbrus
mountain in a winter resort.  I had my report on the masses of galaxies on
January 29, 1974. My message was: {\em since the data suggest that all
giant galaxies have massive coronae, dark matter must be the dominating
component in the whole universe}.  In the Winter School prominent Soviet
astrophysicists like Zeldovich, Shklovsky, Novikov and others
participated.  After the talk the atmosphere was as if a bomb had
exploded.  Everybody realised that, if true, this is a discovery of
principal importance. Two questions dominated: What is the physical nature
of the dark matter?  and What is its role in the evolution of the
Universe?

\begin{figure}[ht]
\begin{center}
\begin{minipage}{0.6\textwidth}
\plotone{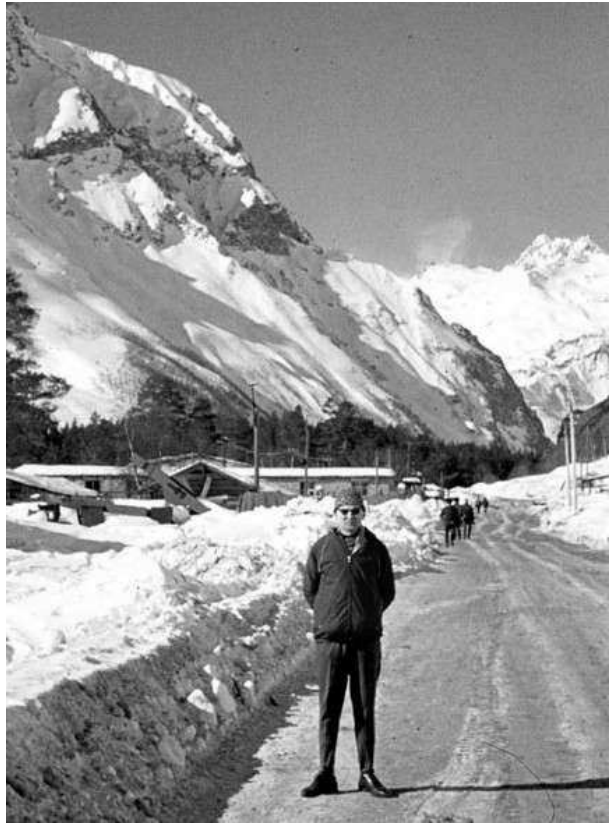}
\end{minipage}
\end{center}
\caption{Author in the Caucasus winter school.}
\label{fig:JE}
\end{figure}

We had to hurry, since large masses of halos were already discussed by
Ostriker and Peebles (1973).  Preliminary results of our analysis were
published in February 1974 by Einasto \etal (1974b) in Astron.  Tsirk.
But Zeldovich insisted that this is not enough: ``Major results must
be be published in major journals''.  Thus a more detailed report was
sent to Nature (Einasto, Kaasik \& Saar 1974) and, for the first time, a
preprint was made and sent to all observatories.  Soon we realised
that it was just in time: Ostriker, Peebles and Yahil (1974) got
similar results using similar arguments; their paper was published
several months after our Nature paper and has a reference to our
preprint.  Soon the first reaction to our results appeared: Burbidge
(1975) formulated difficulties of the dark corona concept.  The main
problem is in the statistical character of the dynamical determination
of the mass of double galaxies.  If companion galaxies used in mass
determination are not real physical companions but random interlopers,
then the mean velocity dispersion reflects random velocities of field
galaxies and no conclusions on the mass distribution around giant
galaxies can been made.  The latter three publications initiated the
dark matter boom.

\begin{figure}[ht]
\plottwo{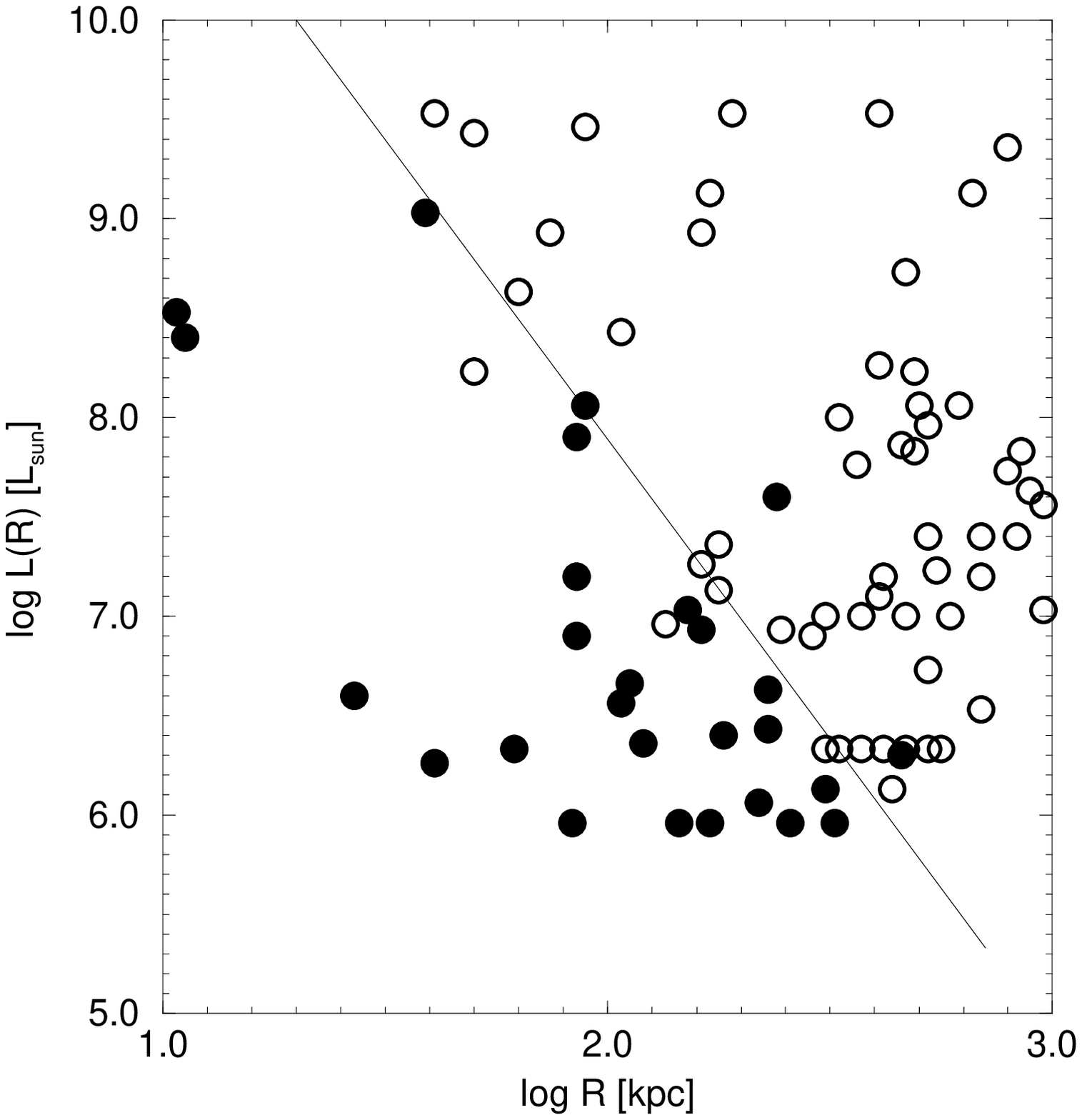}{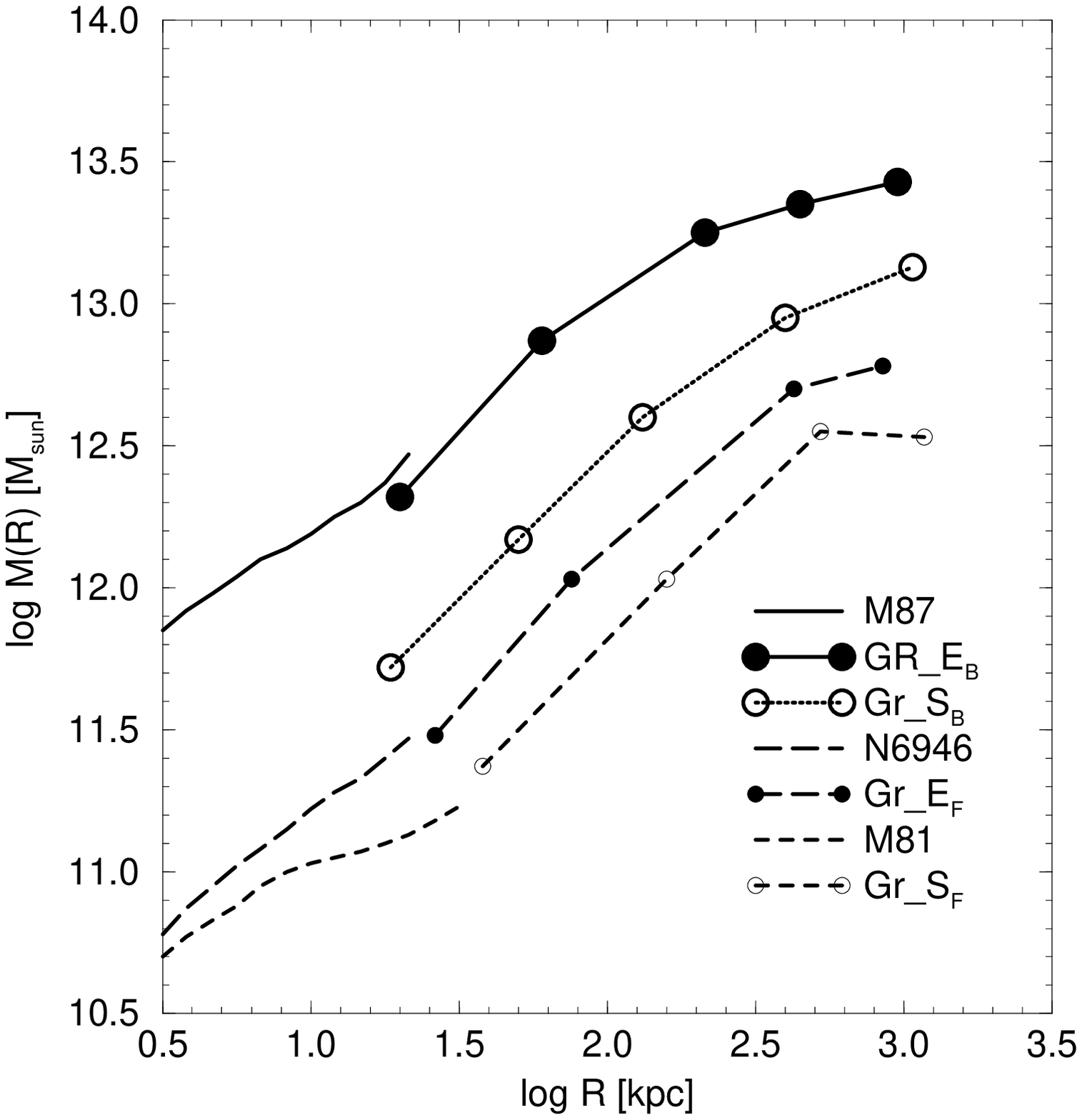}
\caption{Left: Distribution of luminosity of companion galaxies of
different morphology vs. distance from the central galaxy; spiral and
irregular companions are marked with open circles, elliptical
companions with filled circles (Einasto et al. 1974). Right:
Distribution of internal mass in the giant elliptical galaxy M87,
giant spiral galaxy NGC6946, and medium luminous spiral galaxy M81,
compared with mass distribution in groups of galaxies derived from
relative motions of companions of giant and medium bright elliptical
and spiral galaxies (Einasto et al. 1975).
}
\label{fig:dm}
\end{figure}

Difficulties connected with the statistical character of our arguments
were discussed already in the Winter School, thus we started
immediately a study of properties of companion galaxies to find
evidence for some other regularity in the satellite system which
surrounds giant galaxies.  Soon we discovered that companion galaxies
are segregated morphologically: elliptical (non--gaseous) companions
lie close to the primary galaxy whereas spiral and irregular (gaseous)
companions of the same luminosity have larger distances from the
primary galaxy; the distance of the segregation line from the primary
galaxy depends on the luminosity of the primary galaxy (Einasto \etal
1974a, see Figure~5).  This result shows, first of all, that
companions are real members of these systems -- random by--flyers
cannot have such properties.  Second, this result demonstrated that
diffuse matter can have a certain role in the evolution of galaxy
systems.  The role of diffuse matter in galactic coronae was discussed
in detail by Chernin, Einasto \& Saar (1976).  Morphological
properties of companion galaxies can be explained, if we assume that,
at least part of the corona is gaseous.  On the other hand, Komberg \&
Novikov (1975) demonstrated that coronae cannot be fully gaseous.
Thus the nature of coronae remained unclear.  Also we found that
dynamical and morphological properties of primary galaxies are well
correlated with properties of their companions (Einasto \etal 1976c,
see Figure~5).  A further evidence of the large mass of the corona of
our Galaxy came from the study of the dynamics of the Magellanic
Stream (Einasto \etal 1976a). The stellar component of Galactic
coronae was discussed by J\~oeveer (1974a) and Jaaniste \& Saar
(1975). The publication story of the last paper is interesting. First
authors submitted the paper to {\em Astrophysics and Space Science},
but the editor S.B. Pikelner rejected the paper with justification:
``you already have a paper on dark matter'' (by Chernin, Einasto \&
Saar 1976). He could not imagine that in years to come thousands of
papers will be written on this subject.

In January 1975 the first conference on dark matter was held in
Tallinn, Estonia.  The rumor on dark matter had spread around the
astronomical community and, in contrast to conventional local
astronomy conferences, leading Soviet astronomers and physicists
attended.  Main topics were new data for the evidence of the dark
matter, and its possible nature.  Here also statistical arguments
against the dark matter concept were presented by Fessenko.  The next
dark matter discussion was in July 1975 during the Third European
Astronomical Meeting in Tbilisi, Georgia, where a full session was
devoted to the dark matter problem.  The principal discussion was
between our group (Einasto \etal 1976d) and Materne \& Tammann (1976).
Historically, this was the first well documented discussion between
supporters and opponents of the galactic dark matter concept.  Tammann
ignored our arguments, as well as arguments by Zwicky, Kahn and
Woltjer. Using his own statistical data he concluded that systems of
galaxies are stable with conventional masses. It was clear that by
sole dispute it is not possible to solve the problem -- new data were
needed. These data were supplied by Vera Rubin and her collaborators
in their analysis of rotation curves of galaxies (Rubin, Ford \&
Thonnard 1978, 1980, see also a review by Rubin 1987).  These studies
indicated that practically all spiral galaxies have flat rotation
curves as expected in the presence of dark coronae (halos).

Dark matter problems were also discussed during the IAU General
Assembly in Grenoble, 1976.  Here arguments for the non--stellar
nature of dark coronae were again presented (Einasto, J\~oeveer \&
Kaasik 1976b). I remember that after my talk Ivan King quietly said
from the audience {\em ``Perhaps really there are two halos of
galaxies''} (stellar halo and non--stellar corona). The concept of the
dark matter was, however, generally accepted by theorists, only after
the need for dark matter in the evolution of the Universe was
clarified.  Rees (1977) noticed that neutrinos can be considered as
dark matter particles; and Chernin (1981) showed that, if dark matter
is non-baryonic, then this helps to explain the paradox of small
temperature fluctuations of cosmic microwave background radiation.
Density perturbations of non-baryonic dark matter start growing
already during the radiation-dominated era whereas the growth of
baryonic matter is damped by radiation.  If non-baryonic dark matter
dominates dynamically, the total density perturbation can have an
amplitude of the order $10^{-3}$ at the recombination epoch, which is
needed for the formation of the observed structure of the Universe.
This problem was discussed in a conference in Tallinn in April 1981.
Here all prominent Soviet cosmologists and particle physicists
participated (this conference was probably the birth of the
astro--particle physics).  The central problem was the nature of the
dark matter.  In the conference banquet Zeldovich hold an enthusiastic
speech: {\em ``Observers work hard in sleepless nights to collect
data; theorist interpret observations, are often in error, correct
their errors and try again; and there are only very rare moments of
clarification.  Today it is one of such rare moments when we have a
holy feeling of understanding secrets of the Nature.''}  Non-baryonic
dark matter is needed to start structure formation early enough.  This
example illustrates well the attitude of theorists to new
observational discoveries -- the Eddington's test: {\em ``No
experimental result should be believed until confirmed by theory''}
(cited after Turner 1999).  Now, finally, the presence of dark matter
was accepted by leading theorists.

This was, however, not the end of the story.  Soon it was realized
that neutrino--dominated or hot dark matter generates almost no fine
structure of the Universe.  But this is already part of the next
story. 

\section{Large--Scale Structure of the Universe}

\subsection{Zeldovich Question}

After my talk in the Caucasus Winter School in 1972 Zeldovich turned
to me and offered collaboration in the study of the universe.  He was
developing a theory of the formation of galaxies (the pancake theory);
an alternative whirl theory was suggested by Ozernoy, and a third
theory of hierarchical clustering by Peebles.  Zeldovich asked for our
help in solving the question: Can we find some observational evidence
which can be used to discriminate between these theories?

\begin{figure}[ht]
\begin{center}
\begin{minipage}{0.7\textwidth}
\plotone{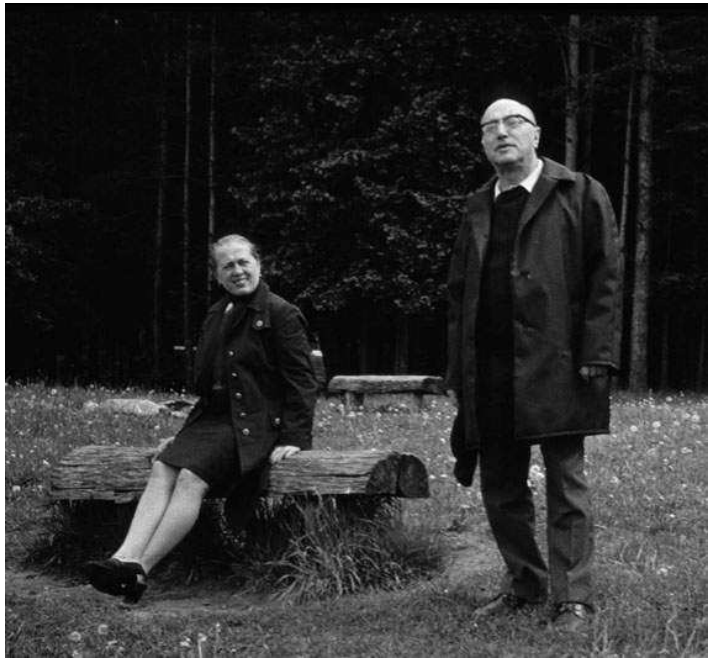}
\end{minipage}
\end{center}
\caption{Yakov Zeldovich with his wife visiting Estonia, late 1970s.}
\label{fig:Zeld}
\end{figure}

So far we had no experience in observational cosmology, our work was
directed to the understanding of the structure of galaxies.  We had
theoretical cosmologists in our group (Enn Saar and Jaak Jaaniste),
but they also did not have experience in observational cosmology.
Thus, initially we had no idea how we can help Zeldovich.  But soon we
remembered our previous experience in the study of galactic
populations: kinematical and structural properties of populations hold
the memory of their previous evolution and formation (Rootsm\"ae 1961,
Eggen, Lynden--Bell \& Sandage 1962).  Random velocities of galaxies
are of the order of several hundred km/s, thus during the whole
lifetime of the Universe galaxies have moved from their place of
origin only about 1~\Mpc\ (we use in this paper the Hubble constant in
units $H_0 = 100~h$ km~s$^{-1}$~Mpc$^{-1}$).  In other words -- if
there exist some regularities in the distribution of galaxies, these
regularities must reflect the conditions in the Universe during the
formation of galaxies.  Actually we had already some first results:
the study of companion galaxies had shown that dwarf galaxies are
located almost solely around giant galaxies and form together with
giant galaxies systems of galaxies.  In other words -- the formation
of galaxies occurs in larger units, not in isolation. 

\begin{figure}[ht]
\begin{center}
\begin{minipage}{0.7\textwidth}
\plotone{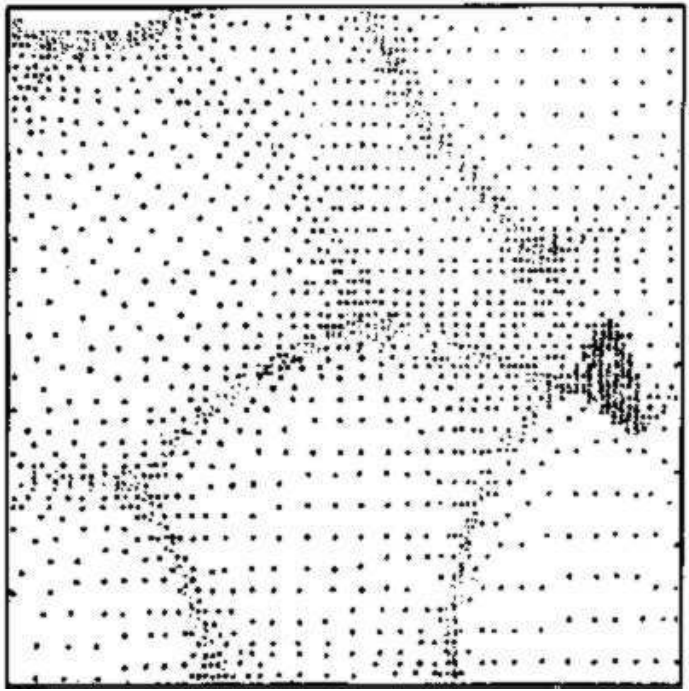}
\end{minipage}
\end{center}
\caption{Distribution of particles in simulations (Doroshkevich,
Shandarin and Novikov 1975).}
\label{fig:model}
\end{figure}

\subsection{Superclusters, Voids and Filaments}

In our work to solve the Zeldovich question we had close collaboration
with his team.  In 1975 Doroshkevich, Shandarin and Novikov obtained
first results of numerical simulations of the evolution of particles
according to the theory of gravitational clustering developed by
Zeldovich (1970).  This was a 2--dimensional simulation with $32\times
32$ particles; a figure with results of these simulations was put on
the wall of the Saar and Jaaniste office (see Figure~7). In this
picture a system of high-- and low--density regions was seen:
high-density regions form a cellular network which surrounds large
under--dense regions.  One of our challenges was to find out whether
the real distribution of galaxies showed some similarity with the
theoretical picture.

Now we had a leading idea how to solve the problem of galaxy
formation: {\em We have to study the distribution of galaxies on larger
scales}.  Both our galactic astronomy and theoretical cosmology groups
participated in this effort.  One approach we used was the study of
the distribution of nearby Zwicky clusters.  Many bright galaxies of
nearby Zwicky clusters had at this time measured redshifts, so we
hoped to determine the distribution of clusters and to find some
regularities there.  To see the distribution better we built in the
office of Saar and Jaaniste a 3--dimensional model from plastic balls.
Some regularity was evident: there were several clusters of Zwicky
clusters -- superclusters, one of them in the Perseus region.  But too
many clusters had no galaxies with measured redshifts, so it was
difficult to get an overall picture.

A different approach was used by Mihkel J\~oeveer.  He used
wedge--diagrams, invented just when we started our study. His trick
was: he made a number of relatively thin wedge diagrams in sequence,
and plotted in the same diagram galaxies, as well as groups and
clusters of galaxies.  In these diagrams a regularity was clearly
seen: {\em isolated galaxies and galaxy systems populated identical
regions, and the space between these regions was empty}.  After this
success the whole group continued the study using wedge--diagrams.
Most attention was directed to the Perseus supercluster, well seen as
a chain of Zwicky and Abell clusters.  To our great surprise and joy
slices with galaxies and clusters were quite similar to the predicted
picture. We made wedge--diagrams for the full sky, in declination and
right ascension (see Figure~8), and a very detailed analysis of the
Perseus supercluster region, where the number of foreground galaxies
was very small (see Figure~9).

\begin{figure}[ht]
\plottwo{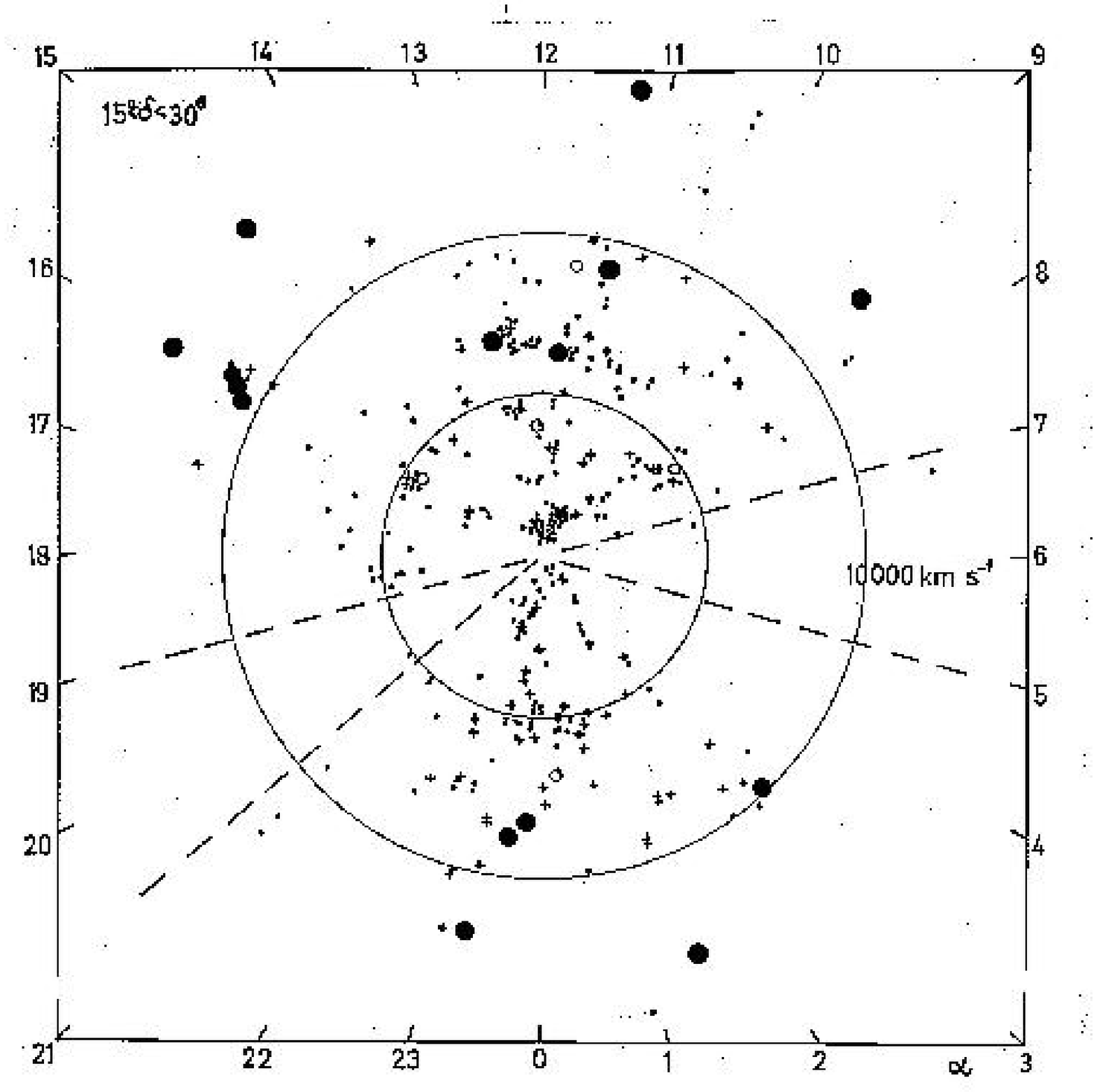}{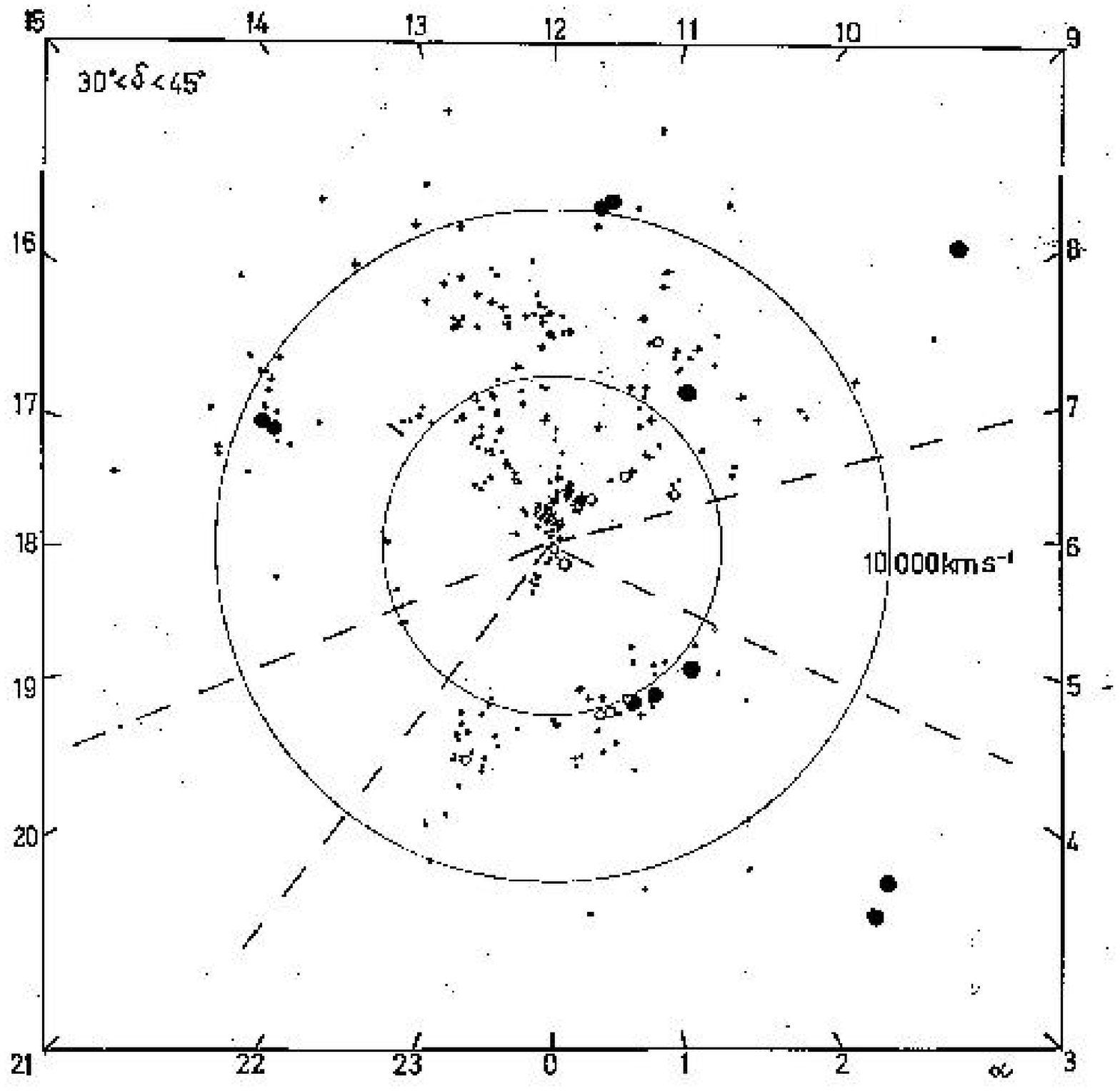}
\caption{Wedge diagrams of galaxies (dots), radio galaxies (crosses),
clusters (filled circles) and groups (open circles) in two declination
zones (J\~oeveer and Einasto 1978). On left panel we see the Local
Supercluster near the center, the Coma and the Hercules superclusters
(at R.A. 12$^h$ and redshift 7000 km/s, and 15$^h$ and redshift 12000
km/s, respectively); on right panel the Perseus Supercluster at
R.A. 1$^h$ and redshift 5000 km/s, and the northern part of the
Hercules supercluster. Also filaments of galaxies joining
superclusters, and voids between superclusters are well
seen. Long-dashed lines indicate the zone of avoidance.
}
\label{fig:wedge}
\end{figure}

\begin{figure}[ht]
\begin{center}
\begin{minipage}{0.9\textwidth}
\plotone{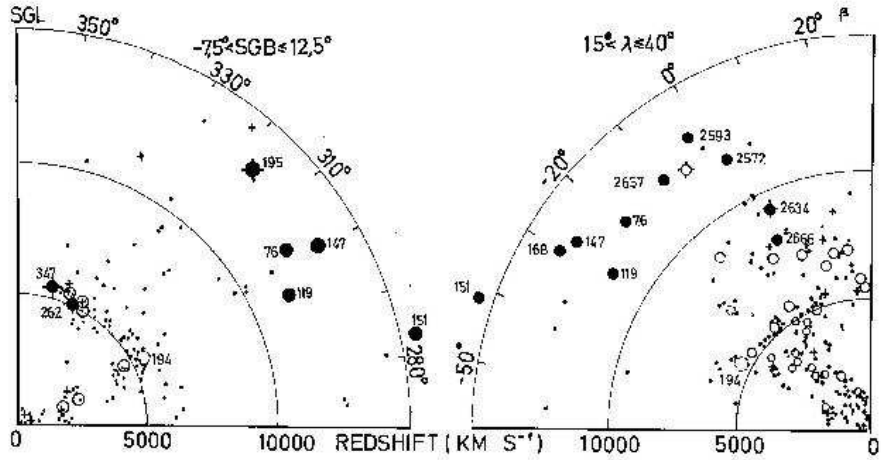}
\end{minipage}
\end{center}
\caption{Wedge diagrams in supergalactic coordinates (left) and in
coordinates parallel to the main chain of the Perseus supercluster
(right) (J\~oeveer and Einasto 1978).  Designations as in Figure~8;
Abell cluster names are indicated.  The presence of galaxy filaments
joining the Perseus and Local superclusters is well seen, as well as
voids in front of and behind the Perseus supercluster.  The last void
has a depth from 5000 to 13000 km/s in redshift.}
\label{fig:pers}
\end{figure}

Already in 1975, after the Tbilisi Meeting, we discussed with
Zeldovich the possibility to organise a real international conference
devoted solely to cosmology.  Due to Soviet bureaucratic system it was
extremely difficult for Soviet astronomers to attend international
conferences in Western countries; thus the only possibility to have a
better contact between Soviet and western cosmologists was to hold the
conference within the Soviet Union.  Zeldovich suggested to hold the
symposium in Tallinn.  After some discussion we decided to devote it
to ``Large Scale Structure of the Universe''.  The symposium was held
in September 1977. The chairman of the organizing committee was
Malcolm Longair. He has spent a long period in Moscow and was well
familiar with the work of Moscow theorists, and Zeldovich fully
trusted him.

Our main results reported during the symposium by J\~oeveer \& Einasto
(1978) were: (1) galaxies, groups and clusters of galaxies are not
randomly distributed but form chains, concentrated in superclusters;
(2) the space between galaxy chains contains almost no galaxies and
form holes (voids) of diameter up to $\approx 70$~\Mpc\ (see
Figure~9); (3) the whole picture of the distribution of galaxies and
clusters resembles cells of a honeycomb, rather close to the picture
predicted by Zeldovich.

The presence of holes (voids) in the distribution of galaxies were
reported also by other groups: by Tully \& Fisher (1978), Tifft \&
Gregory (1978), and Tarenghi \etal (1978) in the Local, Coma and
Hercules superclusters, respectively. A few years later a large void
in Bootes was discovered by Kirshner \etal (1981). Theoretical
interpretation of the observed cellular structure was discussed by
Zeldovich (1978).  As noted by Longair (1978) in his concluding
remarks, {\em the discovery of the filamentary character of the
distribution of galaxies, similar to a lace--tablecloth, and the
overall cellular picture of the large--scale distribution was the most
exciting result presented at this symposium}.  These results
demonstrated that the pancake scenario by Zeldovich (1970) has many
advantages over other rivalling scenarios.  The term ``Large--Scale
Structure of the Universe'' got its present meaning.

\begin{figure}[ht]
\begin{center}
\begin{minipage}{0.8\textwidth}
\plotone{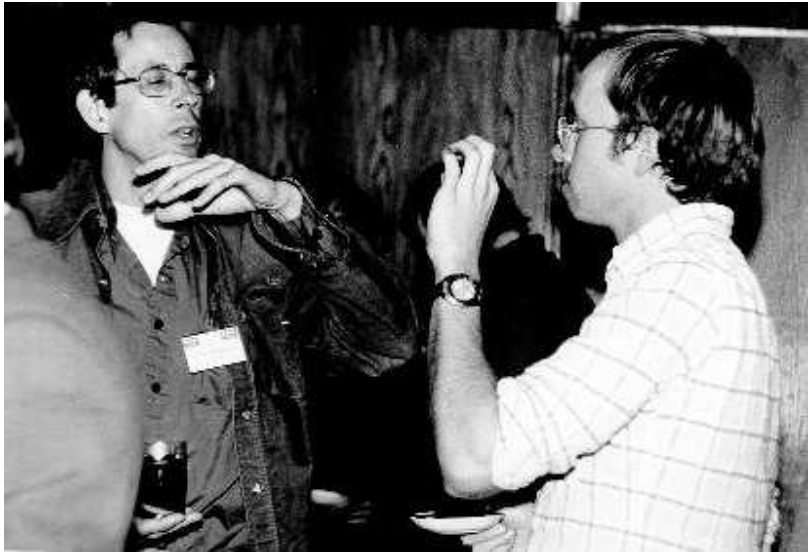}
\end{minipage}
\end{center}
\caption{Jim Peebles explaining  Scott Tremain secrets of structure
formation, IAU Tallinn Symposium 1977}
\label{fig:Peebles}
\end{figure}

A more detailed version of our results was published by J\~oeveer,
Einasto \& Tago (1978).  The next problem to solve was to find some
explanation for the absence of galaxies in voids.  This was done by
Einasto, J\~oeveer \& Saar (1980).  Saar developed an approximate
model of the evolution of density perturbations in under-- and
over--dense regions based on Zeldovich (1970) ideas.  He found that
the matter flows out of under--dense regions and collects in
over--dense regions until it collapses (pancake forming) and forms
galaxies and clusters.  In under--dense regions the density decreases
continuously, but never reaches zero: there must be primordial matter
in voids.  Galaxy formation occurs not everywhere but only after the
matter has collapsed to pancakes.  Originally we believed that
pancakes are 2--dimensional surfaces as predicted by Zeldovich (1970).
To our surprise we did not find evidence for the presence of
wall--like structures between voids -- {\em the dominating structural
element was a chain (filament) of galaxies and clusters}.  Later the
absence of wall--like pancakes and the dominance of filaments was
explained theoretically by Bond, Kofman \& Pogosyan (1996).

Our first 3--dimensional picture of the distribution of galaxies was
based on the Second Reference Catalogue of Galaxies by de Vaucouleurs,
de Vaucouleurs, \& Corwin (1976).  As new redshifts became available
(Sandage \& Tammann 1981, the first Harvard Center for Astrophysics
redshift survey, and ZCAT by Huchra 1981), we made more detailed
analyses of the Local, Coma, Perseus and Hercules superclusters, and
the huge void between these superclusters.  In collaboration with Dick
Miller we prepared several movies where the third dimension was
visualised by rotation of the galaxy or cluster sample (Einasto \&
Miller 1983).  Also quantitative methods were applied to describe the
structure -- correlation function, cluster and percolation analysis,
multiplicity analysis of galaxy systems (Einasto, Klypin \& Shandarin
1983, Einasto \etal 1984).  Our results were summarised in a review
article by Zeldovich, Einasto \& Shandarin (1982) with the following
principal conclusions. Most galaxies and clusters are concentrated in
superclusters and are aligned along strings; strings join in central
rich clusters of galaxies; the whole distribution is cellular.
Large--scale structure of the Universe changes slowly, thus the
present structure reflects the history of its formation and evolution.
The space between galaxy and cluster chains and superclusters is
almost devoid of visible objects; galaxies form only in dense and cold
gas, thus no galaxies will be formed in interiors of voids and
intensive galaxy formation occurs along lines (chains) and knots
(clusters).  Cluster and multiplicity analysis showed that there were
several important differences between models and observations: in all
models there was a population of rarefied particles in voids not
represented as galaxies in the real Universe (compare Figures~7, 8 and
9).  The connectivity of systems was in agreement with observations
only in the pancake model (in contrast to the hierarchical clustering
model). No model yielded a multiplicity function in agreement with
observations: in the real Universe systems of galaxies of all
multiplicity exist, in hierarchical models there are no large
supercluster--type systems, pancake models yield no systems of
intermediate and small multiplicity (there is no fine structure of
small filaments in voids as in the real Universe).

\subsection{Cold Dark Matter, Biasing and Regularity of the Structure}

In late 1981 I was visiting ESO to analyse the large scale structure and
to prepare with Dick Miller movies on the distribution of galaxies. At
this time Oort was preparing a review paper on superclusters (Oort 1983).
To discuss the structure of superclusters with Jan Oort, I had to make a
special arrangement.  Under Soviet rules I had no permission to visit the
Netherlands during this visit, so we agreed with Oort to meet in Bonn, the
German observatory closest to the Netherlands.  After the discussion we
agreed that there exists strong evidence for the formation of galaxies in
chains: the velocity dispersion of groups and galaxies perpendicular to
the chain axis is practically zero; main galaxies in clusters are
elongated along the axis of the chain (as seen in the Perseus chain by
J\~oeveer, Einasto \& Tago 1978 and Einasto, J\~oeveer \& Saar 1980).  
These results were important in several aspects.  First, it was the first
demonstration of biased galaxy formation (no galaxies form in voids).
Secondly, galaxy formation occurs {\em in situ} in chains and clusters.
Third, all previous structure formation scenarios had weak points, and a
new better scenario had to be suggested.  A new scenario was suggested,
among others, by Bond, Szalay \& Turner (1982); here hypothetical
particles like axions, gravitinos or photinos play the role of dark
matter.  Numerical simulations of structure evolution for neutrino and
axion--gravitino--photino--dominated universe were made and analysed by
Melott \etal (1983).  All quantitative characteristics (connectivity of
the structure, multiplicity of galaxy systems, correlation function) of
this new model fit the observational data well.  This model was called
subsequently the Cold Dark Matter (CDM) model, in contrast to the
neutrino--based Hot Dark Matter model. Presently the CDM model with some
modifications is the most accepted model of the structure evolution
(Blumenthal \etal 1984).

The presence of voids and superclusters was generally accepted after
the work of the Harvard group (Geller \& Huchra 1989).  It is a bit
strange that Harvard astronomers initially avoided the term ``void''
and used a new term, ``bubble''; also they coined the terms ``walls''
and ``great walls'' for filaments and superclusters, respectively.

In the 1980s our cosmology group was expanded by 4 young cosmologists Lev
Kofman, Dmitri Pogosyan, Maret Einasto and Mirt Gramann.  Kofman studied
the theory of inflation (with Andrei Linde) and models dominated by
cosmological term (together with Alexei Starobinsky). Following a
suggestion by Enn Saar, Mirt Gramann performed numerical simulations of
models with the cosmological term, initially (in 1984) with a small number
of particles. Soon Tartu Observatory got its first UNIX computer with 2 MB
of core memory; so it was possible to simulate 3--D models with $64^3$
particles and mesh.  One run took about a month -- not too much at those
days.  The main arguments in favour of this model were: direct density
estimates of the mean matter density (including dark matter in voids and
galaxy systems) yield a low value of the density parameter, $\Omega_0
\approx 0.2$; only a model with the cosmological term fits simultaneously
data on the Hubble constant and the age of the Universe. Inflation models
predict that the total density of the Universe is equal to the critical
density which is possible only in the presence of a cosmological constant
(Starobinsky 1982, Kofman \& Starobinsky 1985).  This model was calculated
by Gramann (1988) and was compared with observational data using various
quantitative methods (Einasto, Einasto \& Gramann 1989, Gramann 1990,
Einasto \etal 1991).  All tests suggested that this model fits all
observational data very well.  This LCDM model was probably the first use
of the presently popular model with the cosmological term.  Independent
observational evidence favouring a CDM model with the cosmological term
was found by Efstathiou \etal (1990).

One major issue was the study of physical biasing.  Here we found
that the most important effect is the absence of galaxy formation in
voids and the presence of primordial dark matter here (Einasto,
J\~oeveer \& Saar 1980, Einasto \& Saar 1987).  The presence of an
almost homogeneous population in voids increases the amplitude of the
power spectrum of galaxies in respect to matter; the amplitude shift
is determined by the fraction of matter in the clustered population
associated with galaxies (Einasto \etal 1994a, 1999b).

Kofman, Einasto and Linde (1987) discussed voids (cosmic bubbles) as
remnants from inflation.  Kofman and Shandarin (1988) invented the
adhesion model of the evolution of the large scale structure.  This
model uses simple geometric constructions to calculate, in the first
approximation, the formation and further evolution of the filamentary
web of cosmic structures using the initial density perturbation field.

In the 1990s the main attention of our group was directed to the detailed
analysis of the regularity of the supercluster--void network. Some hints
to the concentration of clusters of galaxies to a dominant plane close to
the supergalactic plane were known already in the 1980s (Einasto \& Miller
1983, Tully \etal 1992).  This issue became topical after the discovery by
Broadhurst \etal (1990) that the distribution of high--density regions may
be quasi-regular or periodic.  Broadhurst \etal found such a regularity in
the direction of the north and south Galactic poles.  Our study confirmed
that this regularity is indeed 3--dimensional -- high--density regions
marked by rich superclusters form a quasi--regular lattice (Einasto \etal
1994b, 1997a,b,c,d). This regularity may be caused by the presence of a
peak or bump in the primordial power spectrum of matter (Einasto \etal
1999a,b,c).  The presence of such feature is predicted in some variants of
the inflation theory (Lesgourgues, Polarski, \& Starobinsky, 1998, Chung
\etal 1999). The issue is still not solved; it is possible that a paradigm
shift is needed here, since no generally accepted theory of structure
formation yields a regular structure.

\section{Epilogue}

In the spring of 1973 there was a conference on galaxies in Tbilisi, and
one evening I walked with Rashid Sunyaev along the Rustaveli avenue
discussing our galaxy models with dark halos.  Sunyaev argued that nobody
will take our results seriously until some American astronomer confirmed
them.  Indeed, our experience has confirmed Sunyaev prediction several
times.  \"Opik's measurement of the distance of the Andromeda galaxy was
accepted only after Hubble's (1925) discovery of Cepheids in M31 and other
galaxies of the Local Group.  Kuzmin's determination of the local mass
density in the Universe remained almost unknown, even after the work by
Gilmore, Wyse and Kuijken.  Our work on the presence of dark matter around
galaxies was noticed after Ostriker, Peebles and Yahil (1974) got similar
results, and Vera Rubin confirmed flat rotation curves of galaxies; the
void--filament structure of the universe was accepted after the work by
Geller and Huchra (1989).  Is this a rule in scientific revolutions?

\acknowledgments 
I thank M. J\~oeveer and E. Saar for fruitful collaboration which has
lasted over 30 years, and H. Andernach for help in improving the
style.  This study was supported by the Estonian Science Foundation
grant 2625.  Spatial distribution of clusters of galaxies is
visualised using computer simulations, see home page of Tartu
Observatory http://www.aai.ee.

\end{document}